\documentclass[twocolumn,aps,superscriptaddress,showpacs,amsmath,amssymb,floatfix,longbibliography]{revtex4-1}
\usepackage{graphicx}
\usepackage{amssymb}
\usepackage{amsmath}
\usepackage{epsfig}
\usepackage{color}
\usepackage{mathtools}
\usepackage[colorlinks,linkcolor=blue,anchorcolor=blue,citecolor=blue,urlcolor=blue]{hyperref}
\usepackage[normalem]{ulem}
\usepackage{diagbox}
\usepackage{mathrsfs}
\usepackage{bigstrut}
\usepackage[section]{placeins}
\usepackage{graphicx}
\usepackage{bm}

\begin{document}
\title{Coupling librational and translational motion of a levitated nanoparticle in an optical cavity}

\author{Shengyan Liu}
\affiliation{Center for Quantum Information, Institute for Interdisciplinary Information Sciences, Tsinghua University, Beijing 100084, China}
\affiliation{Department of Physics, Tsinghua University, Beijing 100084, China}

\author{Tongcang Li}\email{tcli@purdue.edu}
\affiliation{Department of Physics and Astronomy, Purdue University, West Lafayette, IN 47907, USA}
\affiliation{Purdue Quantum Center, Purdue University, West Lafayette, IN 47907, USA}
\affiliation{School of Electrical and Computer Engineering, Purdue University, West Lafayette, IN 47907, USA}
\affiliation{Birck Nanotechnology Center, Purdue University, West Lafayette, IN 47907, USA}

\author{Zhang-qi Yin}\email{yinzhangqi@mail.tsinghua.edu.cn}
\affiliation{Center for Quantum Information, Institute for Interdisciplinary Information Sciences, Tsinghua University, Beijing 100084, China}

\begin{abstract}
An optically levitated nonspherical nanoparticle can exhibit both librational and translational vibrations due to  orientational and translational confinements of the optical tweezer, respectively. Usually, the frequency of its librational mode in a linearly-polarized optical tweezer
 is much larger than the frequency of its translational mode. Because of the frequency mismatch,  the intrinsic coupling between librational and translational modes is very weak in vacuum.
Here we propose a scheme to couple its librational and center-of-mass modes with an optical cavity mode. By adiabatically eliminating the cavity mode, the beam splitter Hamiltonian between librational and  center-of-mass modes can be
realized. We find that high-fidelity quantum state transfer between the librational and translational modes can be achieved with practical parameters. Our work may find applications in sympathetic cooling of multiple modes and quantum information processing.
\end{abstract}

\pacs{******} 

\maketitle

\section{Introduction}
Quantum optomechanics is a rapidly developing field that deals with the interaction between an optical field and the mechanical motion of an object \cite{ASP2014,PZ2012}.  In the last decade, there were a lot of studies on the interaction between the light and the center-of-mass motion of a mechanical oscillator. Quantum ground cooling of mechanical oscillators has been realized \cite{Connell2010,Chan2011}. The study of
optomechanics has many applications in macroscopic quantum mechanics \cite{Chen2013,Yin2017}, precise measurements \cite{Teufel2009}, and quantum information processing \cite{Yin2015,Li2013}.

An optically levitated dielectric nanoparticle in vacuum can have an ultra-high mechanical Q $>10^{9}$ \cite{Chang2010,Romero2010,Li2011,Jain2016}. Therefore, it can be used for ultra-sensitive force detection \cite{Ranjit2016}, searching for hypothetical millicharged particles and dark energy interactions \cite{Rider2016,Moore2014}, and testing the boundary between quantum and classical mechanics \cite{Romero2011,Yin2013}. A levitated nanoparticle has 6 degrees of freedom: three translational modes and three rotational modes \cite{Shi2016}.
If its orientation is confined by the optical tweezer, it will exhibit libration (Such motion was called ``torsional vibration'' in Ref. \cite{Hoang2016,Shi2013}, and ``rotation'' in Ref. \cite{Stickler2016, Kuhn2016}. Several recent papers called it ``libration'' \cite{Nagornykh2016,Zhong2017}, which may be a better term as it is similar to the libration of a molecule in an external field).
The librational  mode of an optically levitated nonspherical nanoparticle has been observed recently \cite{Hoang2016,Kuhn2016}.
Both translational motion and libration of a nanoparticle could be coupled with light and cooled towards quantum ground state by a cavity mode \cite{Shi2016}.
The librational mode frequency could be
one order of magnitude higher than the frequency of a translational mode \cite{Hoang2016}. The coupling between the librational mode and the cavity mode can also be larger than the coupling
between the translational mode and the cavity \cite{Hoang2016}. Therefore, it requires  less cooling laser power to cool the librational mode to the
quantum regime than to cool the translational mode \cite{Hoang2016,Marquardt2007,Wilson2007,Stickler2016}.

In an optical trap in vacuum, the $6$ motional degrees of freedom of a nanoparticle are uncoupled with each other when they are near ground state.
It would be interesting to study how to induce strong coupling between them. Such coupling will have several applications. For example, we may use one of these modes to synthetically cool  other modes.
It is also useful for quantum information, as we may use all $6$ motional modes to store quantum bits, and realize quantum processes such as controlled gates.
By dynamically tuning the polarization orientation of a trapping laser, it was found that two different translational modes could be coupled with each other \cite{Frimmer2016}. In this way, one translational mode was synthetically cooled by coupling it to another translational mode, which was feedback cooled.
It has been proposed to couple translational and rotational  motion of a  sphere with a  spot painted on its surface by a continuous joint measurement of two motional modes \cite{Ralph2016}.
However, a coherent way to couple rotational and translational motion of a nanoparticle is still lacking.

In this paper, we propose a scheme to realize  strong coupling between librational and translational modes of a levitated nanoparticle. We consider an optically trapped nano-particle
that resides in an optical cavity. Both its translational and librational modes couple with the cavity mode.
We discuss the effects of cavity decay, and find that  high-fidelity quantum state transfer could be realized under realistic experimental conditions. We also find that two-mode
squeezing Hamiltonian between librational and translational modes could be realized by adjusting the detunings of driving lasers.

\section{The model}

\begin{figure}[htbp]
\centering
\includegraphics[width=0.45\textwidth]{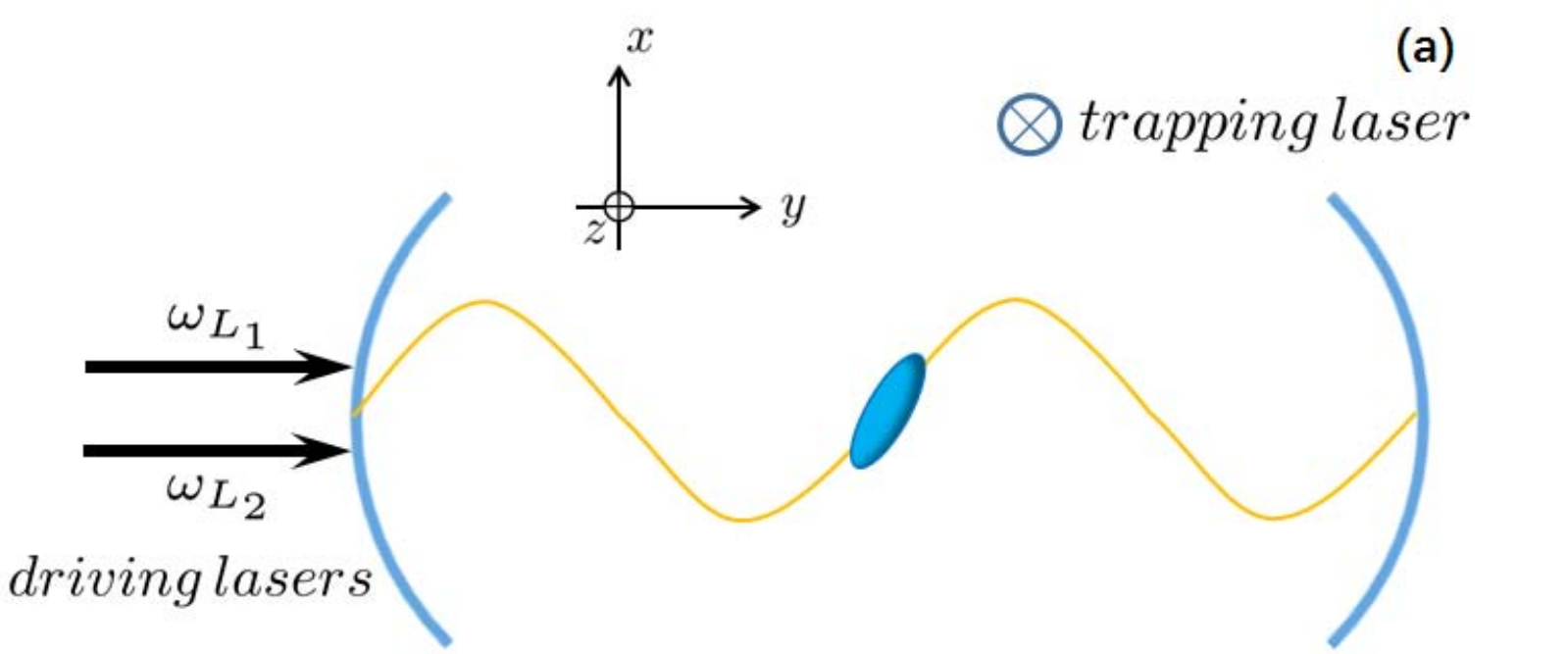}
\includegraphics[width=0.20\textwidth]{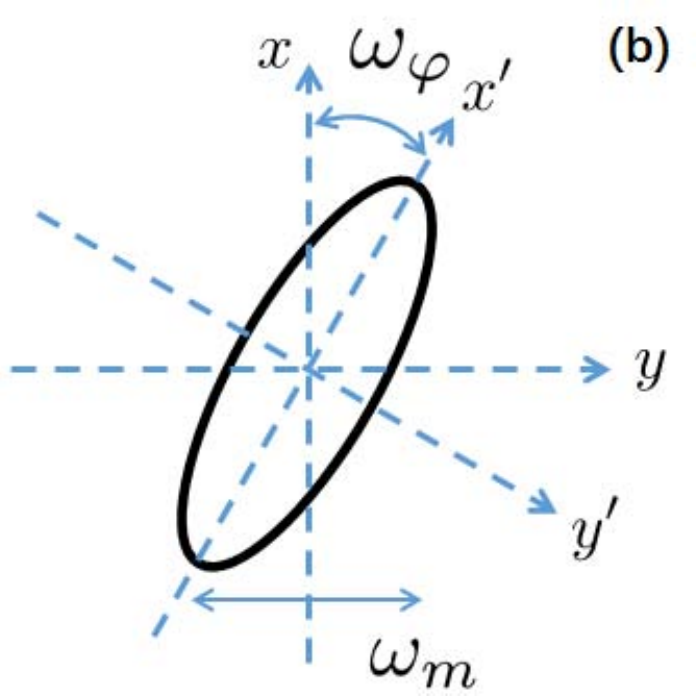}
\caption{(a) The nanodiamond is levitated by the trapping laser and is placed in a cavity. The trapping laser propagates along the z-axis. The cavity is driven by two lasers of frequencies ${\omega _{{L_1}}}$ and ${\omega _{{L_2}}}$ (b) Details of the nanoparticle: $\left( {x,y,z} \right)$ is the coordinate system of the cavity. The  x-axis aligns with the polarization direction of the trapping laser. $\left( {x',y',z'} \right)$ is the coordinate fixed on the nanoparticle. The x'-axis aligns with the long axis of the nanoparticle.}
\label{label}
\end{figure}

As shown in Fig. 1, we consider a system that contains an optical cavity and an ellipsoidal nanoparticle levitated by a trapping laser. The trapping laser is linearly polarized.
Therefore, both location and direction of the nanoparticle are fixed \cite{Hoang2016}.
The nanoparticle has translational mode $b$ with frequency ${\omega _ m }$ and librational mode $c$ with frequency ${\omega _\varphi }$. They are both coupled to the cavity mode $a$. The frequency of the mode $c$ is usually much larger than the frequency of the mode $b$.
The optical mode is driven by two lasers of frequencies ${\omega _{{L_1}}}$ and ${\omega _{{L_2}}}$.
The Hamiltonian of the system can be divided into three parts: ${H_E}$, ${H_I}$ and ${H_D}$. Such that
\begin{equation}\label{eq:Hamiltonian}
H = {H_E} + {H_I} + {H_D}
\end{equation}
where
\begin{equation}
{H_E} = \hbar {\omega _{\rm{0}}}{a^\dag }a + \hbar {\omega _m}{b^\dag }b + \hbar {\omega _\varphi }{c^\dag }c
\end{equation}
\begin{equation}
{H_I} = \hbar {g_{ab}}{a^\dag }a\left( {{b^\dag } + b} \right) + \hbar {g_{ac}}{a^\dag }a\left( {{c^\dag } + c} \right)
\end{equation}
\begin{equation}
{H_D} = \frac{{h{\Omega _{\rm{1}}}}}{2}\left( {{a^\dag }{e^{ - i{\omega _{{L_{\rm{1}}}}}t}} + a{e^{i{\omega _{{L_{\rm{1}}}}}t}}} \right){\rm{ + }}\frac{{h{\Omega _{\rm{2}}}}}{2}\left( {{a^\dag }{e^{ - i{\omega _{{L_2}}}t}} + a{e^{i{\omega _{{L_2}}}t}}} \right)
\end{equation}
Here $H_E$ is the energy term of translational mode $b$, librational mode $c$ and cavity mode $a$. $H_I$ describes the couplings between cavity mode $a$
and two mechanical modes $b$ and $c$. The coupling rates ${g_{ab}}$ and ${g_{ac}}$ are small, but they can be amplified by the driving lasers $H_D$. We will
discuss how to derive the effective Hamiltonian between $a$ and $b$, $c$ modes in the next section.

\section{The effective Hamiltonian}
We firstly consider an ideal system without decay. In order to get the effective Hamiltonian between the cavity mode $a$ and mechanical modes $b$ and $c$, we first give the Heisenberg equation corresponding to (1).
\begin{equation}\label{eq:Heisenberg}
\begin{split}
\dot a =& - i {\omega _0}a - i {g_{ab}}a\left( {{b^\dag } + b} \right) - i{g_{ac}}a\left( {{c^\dag } + c} \right) \\
&- i\frac{{ {\Omega _1}}}{2}{e^{ - i{\omega _{{L_1}}}t}} - i\frac{{ {\Omega _2}}}{2}{e^{ - i{\omega _{{L_2}}}t}}.
\end{split}
\end{equation}
To deal with it, we make a semi-classical ansatz
\begin{equation}\label{eq:ansatz}
a = {a_0}\left( t \right) + {\alpha _1}\left( t \right){e^{ - i{\omega _{{L_{\rm{1}}}}}t}} + {\alpha _2}\left( t \right){e^{ - i{\omega _{{L_2}}}t}}
\end{equation}
where ${\alpha _1}$ and ${\alpha _2}$ are the classical amplitudes of mode $a$ with frequencies $\omega_{L1}$ and $\omega_{L2}$, and $a_0$ is the quantum fluctuation operator.

Inserting \eqref{eq:ansatz} into \eqref{eq:Heisenberg}, we get the equation for the classical amplitudes $\alpha _1$ and $\alpha _2$
\begin{equation}
\begin{array}{l}
 - i{\omega _{{L_{\rm{1}}}}}\alpha _1 {e^{ - i{\omega _{{L_{\rm{1}}}}}t}} - i{\omega _{{L_2}}}\alpha _2 {e^{ - i{\omega _{{L_2}}}t}} + \dot \alpha _1 {e^{ - i{\omega _{{L_{\rm{1}}}}}t}} + \dot \alpha _2 {e^{ - i{\omega _{{L_2}}}t}} = \\
 - i{\omega _0}\alpha _1 {e^{ - i{\omega _{{L_{\rm{1}}}}}t}} - i{\omega _0}\alpha _2 {e^{ - i{\omega _{{L_2}}}t}} - i\frac{{{\Omega _1}}}{2}{e^{ - i{\omega _{{L_1}}}t}} - i\frac{{{\Omega _2}}}{2}{e^{ - i{\omega _{{L_2}}}t}}.
\end{array}
\end{equation}
As $\alpha _1$ and $\alpha _2$ have different frequencies, we have equations for each of them
\begin{equation}
\begin{aligned}
\dot \alpha _1 = &- i {\omega _0}\alpha _1 + i{\omega _{{L_{\rm{1}}}}}\alpha _1 - i\frac{{ {\Omega _1}}}{2}\\
\dot \alpha _2 = &- i {\omega _0}\alpha _2 + i{\omega _{{L_2}}}\alpha _2 - i\frac{{ {\Omega _2}}}{2}
\end{aligned}
\end{equation}
So we can get their classical steady state amplitude ($\dot \alpha_1  = \dot \alpha _2  = 0$): $\alpha _1 = \frac{{{\Omega _1}}}{{2{\Delta _1}}}$ and $\alpha _2 = \frac{{{\Omega _2}}}{{2{\Delta _2}}}$, where ${\Delta _1} = {\omega _{{L_1}}} - {\omega _0}$, ${\Delta _2} = {\omega _{{L_2}}} - {\omega _0}$. So we get
\begin{equation}\label{eq:steadya}
a = {a_0}{e^{ - i{\omega _0}t}} + \frac{{{\Omega _1}}}{{2{\Delta _1}}}{e^{ - i{\omega _{{L_1}}}t}} + \frac{{{\Omega _2}}}{{2{\Delta _2}}}{e^{ - i{\omega _{{L_2}}}t}}
\end{equation}
We can derive steady-state displacements $\beta$ and $\gamma$ for $b$ and $c$ in the same way.
\begin{equation}\label{eq:steadybc}
\begin{aligned}
b = {b_0} + \beta \\
c = {c_0} + \gamma
\end{aligned}
\end{equation}
Where $\beta  =  - {g_{ab}}{{\left( {\alpha _1^2 + \alpha _2^2} \right)} \mathord{\left/
 {\vphantom {{\left( {\alpha _1^2 + \alpha _2^2} \right)} {{\omega _m}}}} \right.
 \kern-\nulldelimiterspace} {{\omega _m}}}$, $\gamma  =  - {g_{ac}}{{\left( {\alpha _1^2 + \alpha _2^2} \right)} \mathord{\left/
 {\vphantom {{\left( {\alpha _1^2 + \alpha _2^2} \right)} {{\omega _\varphi }}}} \right.
 \kern-\nulldelimiterspace} {{\omega _\varphi }}}$. We substitute \eqref{eq:ansatz} and \eqref{eq:steadybc} into Hamiltonian and in the rotating frame with $U = {e^{ - iH_0t/\hbar}}$, where ${H_0} = \hbar {\omega _0}a_0^\dag {a_0} + \hbar {\omega _m}b_0^\dag {b_0} + \hbar {\omega _\varphi }c_0^\dag {c_0}$. The Hamiltonian ${H_{RW}} = {U^\dag }(H - H_0)U$. By tuning the lasers detunings, we can get different Hamiltonian between mechanical modes and the cavity mode. The different tasks such as quantum state transfer and entanglement generating can be realized. For example, if the driving lasers fulfill ${\Delta _1} = -{\omega _m}$ , ${\Delta _2} = -{\omega _\varphi }$, we can neglect fast oscillation terms. The effective Hamiltonian reads
\begin{equation}
{H_{RW}} = \hbar {g_{ab}}\alpha _1 \left( {a_0^\dag b_0 + {a_0}{b_0^\dag }} \right) + \hbar {g_{ac}}\alpha _2 \left( {a_0^\dag c_0 + {a_0}{c_0^\dag }} \right).
\end{equation}
The perfect quantum state transfer between the translational and the librational modes requires $|g_{ab} \alpha _1 |= |g_{ac} \alpha _2|=G$. If we initialize the system as $\left| {{\psi _a}\left( {t = 0} \right)} \right\rangle \left| {{\psi _b}\left( {t = 0} \right)} \right\rangle \left| {{\psi _c}\left( {t = 0} \right)} \right\rangle  = \left| 0 \right\rangle \left| 0 \right\rangle \left| 1 \right\rangle$, we can get
\begin{equation}
\begin{split}
\left| {{\psi _a}{\psi _b}{\psi _c}\left( t \right)} \right\rangle  =& \frac{1}{2}\left( {1 + \cos \sqrt 2 Gt} \right)\left| {001} \right\rangle  \\
&- \frac{1}{2}\left( {1 - \cos \sqrt 2 Gt} \right)\left| {010} \right\rangle  \\
&- \frac{{i\sqrt 2 }}{2}\sin \sqrt 2 Gt\left| {100} \right\rangle.
\end{split}
\end{equation}
\noindent If we let $t = \frac{\pi }{{\sqrt 2 G}}$, we can transfer a state from librational mode to translation mode (vice versa).

If we set ${\Delta _1}{\rm{ + }}{\omega _m}{\rm{ = }}{\Delta _{\rm{2}}}{\rm{ + }}{\omega _\varphi } = \delta$,
and in the large detuning limit $\delta \gg |g_{ab} \alpha _1 |, |g_{ac} \alpha _2|$, the cavity mode can be adiabatically eliminated \cite{James2007}. 
Here we including all fast rotating terms, both rotating wave and anti-rotating wave. If the cavity mode $a_0$ is initially in the vacuum state, the effective Hamiltonian is
\begin{equation}
{H_{eff}} = \hbar {G_1}b_0^\dag {b_0} + \hbar {G_2}c_0^\dag {c_0} + \hbar {G_3}\left( {b_0^\dag {c_0} + {b_0}c_0^\dag } \right)
\end{equation}
where
\begin{equation}
{G_1} = \frac{{\alpha _1^2g_{ab}^2}}{{{\Delta _1} + {\omega _m}}} + \frac{{\alpha _1^2g_{ab}^2}}{{{\Delta _1} - {\omega _m}}} + \frac{{\alpha _2^2g_{ab}^2}}{{{\Delta _2} + {\omega _m}}} + \frac{{\alpha _2^2g_{ab}^2}}{{{\Delta _2} - {\omega _m}}}
\end{equation}
\begin{equation}
G_2=\frac{{\alpha _2^2g_{ac}^2}}{{{\Delta _2} + {\omega _\varphi }}} + \frac{{\alpha _2^2g_{ac}^2}}{{{\Delta _2} - {\omega _\varphi }}} + \frac{{\alpha _1^2g_{ac}^2}}{{{\Delta _1} + {\omega _\varphi }}} + \frac{{\alpha _1^2g_{ac}^2}}{{{\Delta _1} - {\omega _\varphi }}}
\end{equation}.

\begin{equation}
G_3=\left( {\frac{{{\alpha _{\rm{1}}}{\alpha _2}{g_{ab}}{g_{ac}}}}{{{\Delta _1} + {\omega _m}}} + \frac{{{\alpha _{\rm{1}}}{\alpha _2}{g_{ab}}{g_{ac}}}}{{{\Delta _1} - {\omega _\varphi }}}} \right)
\end{equation}

If $G_1=G_2$ (we will provide workable parameters in next section), and we take the initial state as $\left| {{\psi _b}\left( {t = 0} \right)} \right\rangle \left| {{\psi _c}\left( {t = 0} \right)} \right\rangle = \left| 0 \right\rangle \left| 1 \right\rangle$, we can get
\begin{equation}
\begin{split}
\left| {{\psi _b}\left( t \right)} \right\rangle \left| {{\psi _c}\left( t \right)} \right\rangle  &= \frac{1}{2}\left( {{e^{ - i\left( {{G_1} + {G_3}} \right)t}} + {e^{ - i\left( {{G_1} - {G_3}} \right)t}}} \right)\left| 0 \right\rangle \left| 1 \right\rangle  \\
&+ \frac{1}{2}\left( {{e^{ - i\left( {{G_1} + {G_3}} \right)t}} - {e^{ - i\left( {{G_1} - {G_3}} \right)t}}} \right)\left| 1 \right\rangle \left| 0 \right\rangle .
\end{split}
\end{equation}
\noindent In the lab reference frame, we have
\begin{equation}
\begin{split}
\left| {{\psi _b}\left( t \right)} \right\rangle \left| {{\psi _c}\left( t \right)} \right\rangle  =& \frac{1}{2}{e^{ - i{\omega _\varphi }t}}\left( {{e^{ - i\left( {{G_1} + {G_3}} \right)t}} + {e^{ - i\left( {{G_1} - {G_3}} \right)t}}} \right)\left| 0 \right\rangle \left| 1 \right\rangle \\
 +& \frac{1}{2}{e^{ - i{\omega _m}t}}\left( {{e^{ - i\left( {{G_1} + {G_3}} \right)t}} - {e^{ - i\left( {{G_1} - {G_3}} \right)t}}} \right)\left| 1 \right\rangle \left| 0 \right\rangle .
\end{split}
\end{equation}
\noindent If we let $t = \frac{\pi }{{2G_3^2}}$ , we can transfer a state from librational mode to translational mode (vice versa).

 We can also choose ${\Delta _1}{\rm{ - }}{\omega _m}{\rm{ = }}{\Delta _{\rm{2}}}{\rm{ - }}{\omega _\varphi } =   \delta $. In the limit $\delta \gg G$, we can adiabatically eliminate the cavity mode, and get a two-mode-squeezing effective Hamiltonian \cite{Yin2009,Li2013}
\begin{equation}
{H_{RW}} = \hbar {G'_1}\left( {b_0^\dag {b_0} + c_0^\dag {c_0}} \right) + \hbar {G'_3}\left( {b_0^\dag c_0^\dag  + {b_0}{c_0}} \right).
\end{equation}
which could be used for generating entanglement between modes $b_0$ and $c_0$.

\section{Experimental feasibility and dissipation effects}
In this section, we will provide the feasible parameters in experiment and consider the effect of dissipations. In our scheme, the steady-state amplitudes $\alpha _1$ and $\alpha _2$ are in the order of ${10^4}$ to ${10^5}$. Therefore, the strength of linear couplings between cavity mode and the mechanical modes are enhanced by ${10^4}$ to ${10^5}$ times. The photon number fluctuation is in the order of $\sqrt{\alpha_{1,2}} \sim {10^2}$,  which is relating with non-linear coupling between the cavity and the mechanical modes. Therefore, the linear coupling strength is ${10^2}$ times larger than the non-linear coupling strength. The effect of the photon number fluctuation is negligible in our scheme.

In experiments, the dissipation by cavity mode and mechanical modes decay is inevitable. However, in high vacuum, the mechanical decay rates are much less than the cavity decay rate \cite{Hoang2016,Stickler2016a,Zhong2016}.  Therefore, we only need to consider the cavity decay effect. Considering the dissipation, the steady-amplitudes will change and we can derive them by adding a term of $- i\hbar \frac{\kappa }{2}{a^\dag }a$ into Hamiltonian \eqref{eq:Hamiltonian}. Following the same procedure mentioned above, we can get
\begin{equation}
\begin{split}
{\alpha _1} = \frac{{{\Omega _1}}}{{2\left( {{\Delta _1} + i\frac{\kappa }{2}} \right)}}\\
{\alpha _2} = \frac{{{\Omega _2}}}{{2\left( {{\Delta _2} + i\frac{\kappa }{2}} \right)}}
\end{split}
\end{equation}
And in order to maintain the form of the Hamiltonian, we should do the transformation ${b_0} \to {{{\alpha _1}{b_0}} \mathord{\left/
 {\vphantom {{{\alpha _1}{b_0}} {\left| {{\alpha _1}} \right|}}} \right.
 \kern-\nulldelimiterspace} {\left| {{\alpha _1}} \right|}},b_0^\dag  \to {{\alpha _1^*b_0^\dag } \mathord{\left/
 {\vphantom {{\alpha _1^*b_0^\dag } {\left| {{\alpha _1}} \right|}}} \right.
 \kern-\nulldelimiterspace} {\left| {{\alpha _1}} \right|}}$ and ${c_0} \to {{{\alpha _2}{c_0}} \mathord{\left/
 {\vphantom {{{\alpha _2}{c_0}} {\left| {{\alpha _2}} \right|}}} \right.
 \kern-\nulldelimiterspace} {\left| {{\alpha _2}} \right|}},c_0^\dag  \to {{\alpha _2^*c_0^\dag } \mathord{\left/
 {\vphantom {{\alpha _2^*c_0^\dag } {\left| {{\alpha _2}} \right|}}} \right.
 \kern-\nulldelimiterspace} {\left| {{\alpha _2}} \right|}}$.
Using perturbation theory \cite{Chang2010,Buck2003}, we can obtain the coupling constants in the same way with Ref. \cite{Hoang2016}. If we restrict the librational motion of the long axis of the nanoparticle in the plane $xOy$, we get
\begin{equation}
\begin{split}
{g_{ab}} =& \sqrt {\frac{\hbar }{{2M{\omega _m}}}} \frac{{32{\pi ^2}c{e^{ - \frac{{4\pi \left( {{x^2} + {z^2}} \right)}}{{\lambda L}}}}\cos ky\sin ky}}{{{\varepsilon _0}{\lambda ^3}{L^2}}} \\
& \cdot \left( {{s _2} + {{\cos }^2}\varphi \left( {{s _1} - {s _2}} \right)} \right),
\end{split}
\end{equation}
\begin{equation}
{g_{ac}} = \sqrt {\frac{\hbar }{{2I{\omega _\varphi }}}} \frac{{8\pi c{e^{ - \frac{{4\pi \left( {{x^2} + {z^2}} \right)}}{{\lambda L}}}}{{\cos }^2}ky}}{{{\varepsilon _0}{\lambda ^2}{L^2}}}\left( {{s _1} - {s _2}} \right)\sin 2\varphi .
\end{equation}
Here $L$ is the length of the cavity, $\lambda$ and $k$ is the wavelength and wavenumber of cavity mode. $M$ and $I$ is the mass and the moment of inertia of the nanoparticle. ${s _1}$ and ${s _2}$ are the diagonal elements of susceptibility matrix. $\left( {x,y,z,\varphi} \right)$ are the parameters describe the position of the nanoparticle: $\left( {x,y,z} \right)$ are the coordinates of the center of mass (origin is the center of the cavity), $\varphi $ is the angle between the long axis of the nanoparticle and $x$-axis. $x$, $y$, $z$ and $\varphi$ can be changed by adjusting the trapping laser. For example, we choose the angle between the polarization direction of the trapping laser and the y-axis ($\varphi $) as ${45^ \circ }$, the equilibrium position of the center of mass is$\left( {0,\pi/4k,0} \right)$. We can get ${g_{ab}}/2\pi = 0.3056 $ Hz and ${g_{ac}}/2\pi = 0.2189$ Hz. (The parameter of the nano particle we choose: $\rho  = 3500~{{{\rm{kg}}} \mathord{\left/
 {\vphantom {{{\rm{kg}}} {{{\rm{m}}^{\rm{3}}}}}} \right.
 \kern-\nulldelimiterspace} {{{\rm{m}}^{\rm{3}}}}}$, long axis $a = 50$ nm, short axis $b = 25$ nm, ${\varepsilon _r} = 5.7$, waist of the trapping laser ${W_t} = 600$ nm, power of the trapping laser is  $100$ mW, wavelength ${\lambda _{cav}} = 1540$ nm, length of the cavity $L = 10$ mm) And in this situation, ${{{\omega _m}} \mathord{\left/
{\vphantom {{{\omega _m}} {2\pi }}} \right.
\kern-\nulldelimiterspace} {2\pi }} = 247.7$ kHz, ${{{\omega _\varphi }} \mathord{\left/
{\vphantom {{{\omega _\varphi }} {2\pi }}} \right.
\kern-\nulldelimiterspace} {2\pi }} = 2.6$ MHz.
If the finesse of our cavity $\mathcal{F}=10^5$ and we can get ${\kappa \mathord{\left/
{\vphantom {\kappa {2\pi }}} \right.
\kern-\nulldelimiterspace} {2\pi }} = 75.2$ kHz. For example we let ${\delta  \mathord{\left/
 {\vphantom {\delta  {2\pi }}} \right.
 \kern-\nulldelimiterspace} {2\pi }} = 200$ kHz, ${{{\Omega _1}} \mathord{\left/
 {\vphantom {{{\Omega _1}} {{\rm{2}}\pi }}} \right.
 \kern-\nulldelimiterspace} {{\rm{2}}\pi }} = {\rm{2}}{\rm{.66}} \times {10^{{\rm{9}}}}$ Hz, ${{{\Omega _2}} \mathord{\left/
 {\vphantom {{{\Omega _2}} {{\rm{2}}\pi }}} \right.
 \kern-\nulldelimiterspace} {{\rm{2}}\pi }} = 5.0 \times {10^{10}}$ Hz. And we can get ${G_3 \mathord{\left/
{\vphantom {G {2\pi }}} \right.
\kern-\nulldelimiterspace} {2\pi }} = 25$ kHz and time of state transfer $t = 1 \times {10^{ - 5}}$ s thus it's not difficult to realize.

\subsection{Large detuning scheme}
\begin{figure}[htbp]
\centering
\includegraphics[width=0.45\textwidth]{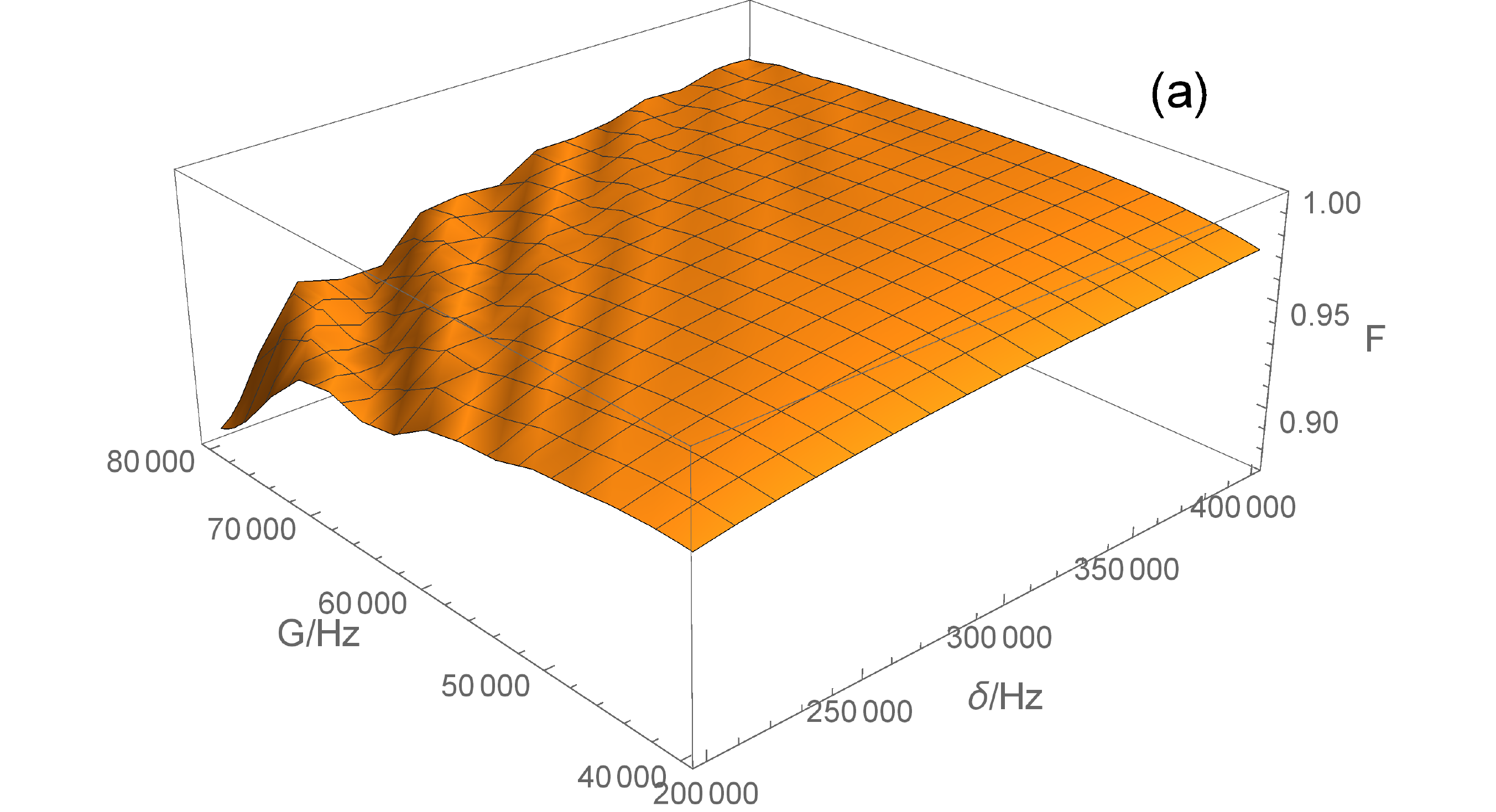}
\includegraphics[width=0.45\textwidth]{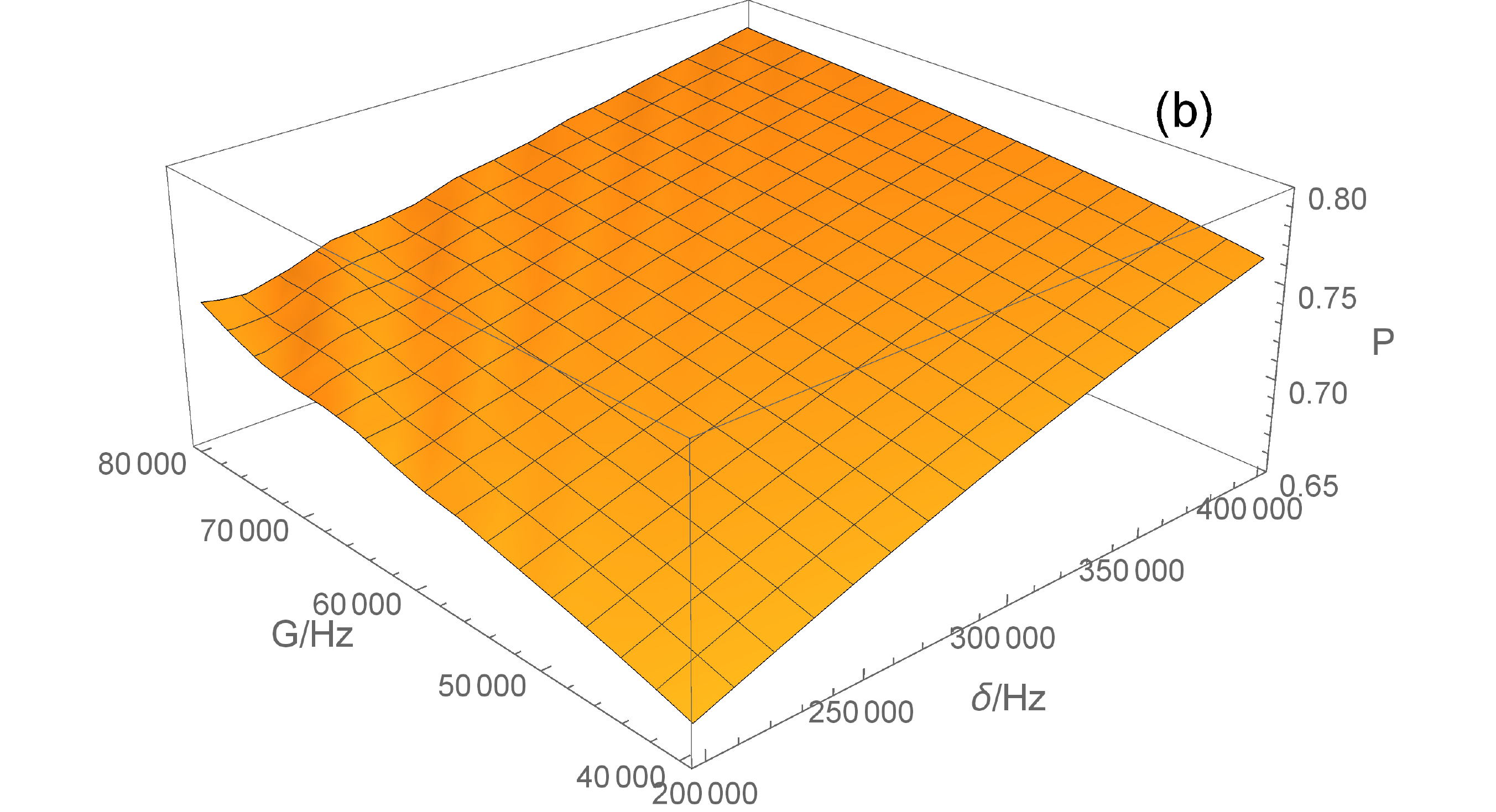}
\caption{{(a) Fidelity of state transfer as a function of $G$ and $\delta$ in the situation of large detuning.  (b) The probability of the system being in this state as a function of $G$ and $\delta$}.}
\label{detuning}
\end{figure}

Under the large detuning condition that ${\Delta _1}{\rm{ + }}{\omega _m}{\rm{ = }}{\Delta _{\rm{2}}}{\rm{ + }}{\omega _\varphi } = \delta \gg G$,  we change the system Hamiltonian to the rotating wave frame, and  neglect the fast rotating terms in $H_{RW}$. In order to deal with the cavity loss effects, here we adopt the conditional Hamiltonian \cite{Plenio1998,Huang2016}. We assume that cavity decay rate is weak. Therfore, we can only consider the situation that system evolves 
without photon leakage.  
Under the condition that no photon is leaking out,  we get the conditional Hamiltonian from quantum trajectory method \cite{Plenio1998}
\begin{equation}
\begin{split}
H =& \hbar G\left( {a_0^\dag b_0{e^{ - i\delta t}} + {a_0}{b_0^\dag }{e^{i\delta t}}} \right) + \hbar G\left( {a_0^\dag c_0{e^{ - i\delta t}} + {a_0}{c_0^\dag }{e^{i\delta t}}} \right) \\
&- i\hbar \frac{\kappa }{2}a_0^\dag {a_0},
\end{split}
\end{equation}
where $\kappa$ is the decay rate of the cavity mode $a$. We can use the above conditional Hamiltonian to calsulate the possibility $P$ of the system evolving without photon leakage. 
Because we suppose that the initial state of the system is${\left| 0 \right\rangle _a}{\left| {01} \right\rangle _{bc}}$, so the subspace only includes 3 basis states:${\left| 0 \right\rangle _a}{\left| {01} \right\rangle _{bc}}$, ${\left| 0 \right\rangle _a}{\left| {10} \right\rangle _{bc}}$ and ${\left| 1 \right\rangle _a}{\left| {00} \right\rangle _{bc}}$. And at any time $t$, the state of the system is
\begin{equation}
\begin{split}
\left| {\psi_d \left( t \right)} \right\rangle =& {C_{d1}}\left( t \right){\left| 0 \right\rangle _a}{\left| {01} \right\rangle _{bc}} + {C_{d2}}\left( t \right){\left| 0 \right\rangle _a}{\left| {10} \right\rangle _{bc}} \\
&+ {C_{d3}}\left( t \right){\left| 1 \right\rangle _a}{\left| {00} \right\rangle _{bc}}
\end{split}
\end{equation}
where
\begin{equation}
{C_{d1}}\left( t \right) = \frac{1}{2} + \frac{{\left( {2\delta + i\kappa + \chi } \right)}}{{4\chi }}{e^{ - i{E_3}t/\hbar }} - \frac{{\left( {2\delta + i\kappa - \chi } \right)}}{{4\chi }}{e^{ - i{E_2}t/\hbar }},
\end{equation}

\begin{equation}
{C_{d2}}\left( t \right) = - \frac{1}{2} + \frac{{\left( {2\delta + i\kappa + \chi } \right)}}{{4\chi }}{e^{ - i{E_3}t/\hbar }} - \frac{{\left( {2\delta + i\kappa - \chi } \right)}}{{4\chi }}{e^{ - i{E_2}t/\hbar }},
\end{equation}

\begin{equation}
\begin{split}
{C_{d3}}\left( t \right) =&  - \frac{{{e^{ - i\delta t}}\left( {2\delta  + i\kappa  + \chi } \right)\left( {2\delta  + i\kappa  - \chi } \right)}}{{16G\chi }} \cdot \\
&\left( {{e^{ - i{E_3}t/\hbar }} - {e^{ - i{E_2}t/\hbar }}} \right)
\end{split}
\end{equation}
here $\chi = \sqrt {4{\delta ^2} + 32{G^2} + 4i\delta \kappa - {\kappa ^2}}$, ${E_2} = \frac{1}{4}\left( { - 2\delta - i\kappa - \chi } \right)$, ${E_3} = \frac{1}{4}\left( { - 2\delta - i\kappa + \chi } \right)$.

 We first normalize the state $\left| \psi_d  \right\rangle$ to calculate the fidelity, we can get $\left| {{\psi _{dn}}} \right\rangle  = {{\left| {{\psi _d}} \right\rangle } \mathord{\left/
 {\vphantom {{\left| {{\psi _d}} \right\rangle } {\sqrt {{{\left| {{C_{d1}}} \right|}^2} + {{\left| {{C_{d2}}} \right|}^2} + {{\left| {{C_{d3}}} \right|}^2}} }}} \right.
 \kern-\nulldelimiterspace} {\sqrt {{{\left| {{C_{d1}}} \right|}^2} + {{\left| {{C_{d2}}} \right|}^2} + {{\left| {{C_{d3}}} \right|}^2}} }}$.
  As shown in Fig. \ref{detuning}(a), we   plot the fidelity $F = \left| {\left\langle {{\psi _{dn}}\left( t \right)\left| {010} \right.} \right\rangle } \right|$ at the time $t = \frac{{\pi \delta }}{{2{G^2} - {{{\kappa ^2}} \mathord{\left/
{\vphantom {{{\kappa ^2}} {16}}} \right.
\kern-\nulldelimiterspace} {16}}}}$ which can be directly derived from the strict solution of the Schr\"{o}dinger Equation and ${\kappa  \mathord{\left/
 {\vphantom {\kappa  {2\pi }}} \right.
 \kern-\nulldelimiterspace} {2\pi }} = 75.2$kHz as a function of $\delta$ and $G$. 
The possibility of the system evolving without photon leakage is $P = {\left| {{C_{d1}}} \right|^2} + {\left| {{C_{d2}}} \right|^2} + {\left| {{C_{d3}}} \right|^2}$. 
It is found that the fidelity could approach to $1$ when $G$ is small and $\delta$ is large. However, at this regime, the effective coupling between two mechanical modes is also pretty small. In Fig. \ref{detuning}(b), we plot $P$ as a function of $\delta$ and $G$ as well. When we choose $\delta = 200$kHz and $G = 50$kHz, the fidelity $F = 0.95$ and the successful possibility $P = 0.68$.

\subsection{Resonant scheme}

\begin{figure}[htbp]
\centering
\includegraphics[width=0.45\textwidth]{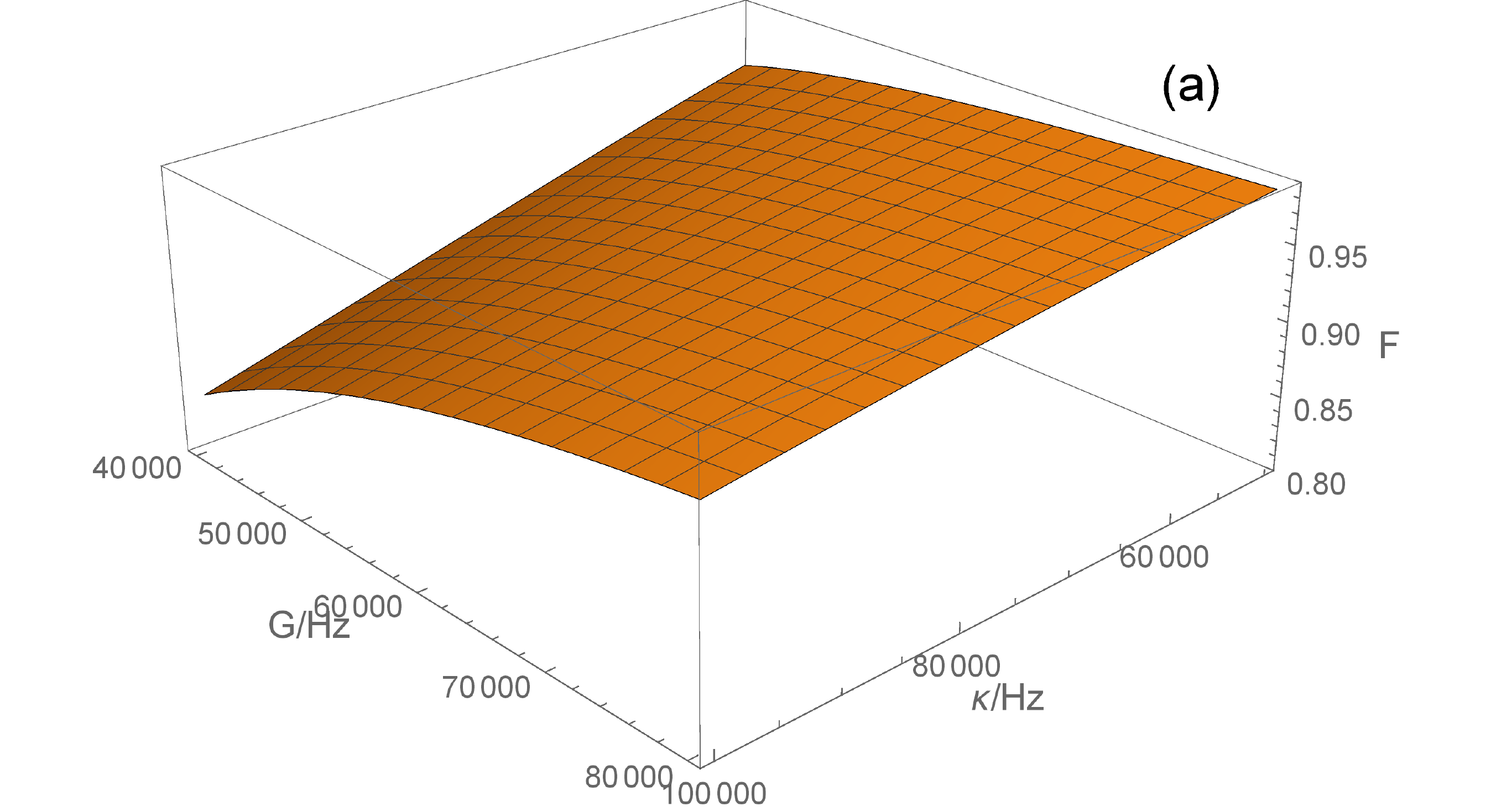}
\includegraphics[width=0.45\textwidth]{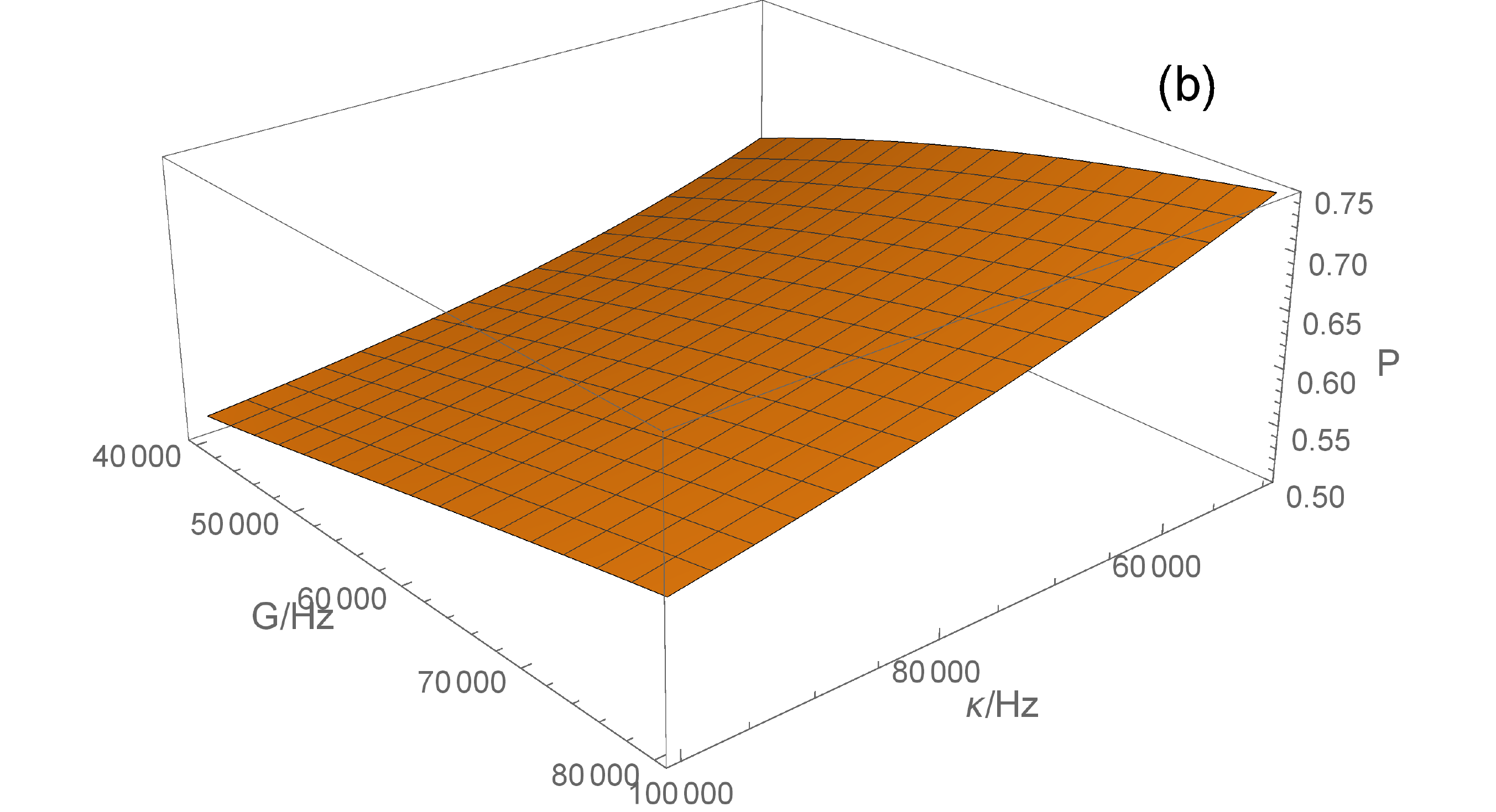}
\caption{(a) Fidelity of state transfer as a function of $G$ and $\kappa$ in the situation of resonance.  (b) The probability of the system being in this state as a function of $G$ and $\kappa$. }
\label{resonant}
\end{figure}

In resonance case, the Hamiltonian reads in
\begin{equation}
H = \hbar G\left( {a_0^\dag b_0 + {a_0}{b_0^\dag }} \right) + \hbar G\left( {a_0^\dag c_0 + {a_0}{c_0^\dag }} \right) - i\hbar\frac{\kappa }{2}a_0^\dag {a_0}
\end{equation}
Because we suppose that the initial state of the system is${\left| 0 \right\rangle _a}{\left| {01} \right\rangle _{bc}}$, the subspace only includes 3 basis states:${\left| 0 \right\rangle _a}{\left| {01} \right\rangle _{bc}}$, ${\left| 0 \right\rangle _a}{\left| {10} \right\rangle _{bc}}$ and ${\left| 1 \right\rangle _a}{\left| {00} \right\rangle _{bc}}$ as well. And at any time $t$, the state of the system is
\begin{equation}
\begin{split}
\left| {{\psi _r}\left( t \right)} \right\rangle  =& {C_{r1}}\left( t \right){\left| 0 \right\rangle _a}{\left| {01} \right\rangle _{bc}} + {C_{r2}}\left( t \right){\left| 0 \right\rangle _a}{\left| {10} \right\rangle _{bc}} \\
&+ {C_{r3}}\left( t \right){\left| 1 \right\rangle _a}{\left| {00} \right\rangle _{bc}}
\end{split}
\end{equation}
and
\begin{equation}
\begin{split}
{C_{r1}}\left( t \right) =& \frac{1}{2} + \frac{1}{2}{e^{ - {{\kappa t} \mathord{\left/
{\vphantom {{\kappa t} 4}} \right.
\kern-\nulldelimiterspace} 4}}}\cos \frac{{\sqrt {32{G^2} - {\kappa ^2}} }}{4}t \\
&+ \frac{\kappa }{{\sqrt {32{G^2} - {\kappa ^2}} }}{e^{ - {{\kappa t} \mathord{\left/
{\vphantom {{\kappa t} 4}} \right.
\kern-\nulldelimiterspace} 4}}}\sin \frac{{\sqrt {32{G^2} - {\kappa ^2}} }}{4}t,
\end{split}
\end{equation}
\begin{equation}
\begin{split}
{C_{r2}}\left( t \right) =& - \frac{1}{2} + \frac{1}{2}{e^{ - {{\kappa t} \mathord{\left/
{\vphantom {{\kappa t} 4}} \right.
\kern-\nulldelimiterspace} 4}}}\cos \frac{{\sqrt {32{G^2} - {\kappa ^2}} }}{4}t \\
&+ \frac{\kappa }{{\sqrt {32{G^2} - {\kappa ^2}} }}{e^{ - {{\kappa t} \mathord{\left/
{\vphantom {{\kappa t} 4}} \right.
\kern-\nulldelimiterspace} 4}}}\sin \frac{{\sqrt {32{G^2} - {\kappa ^2}} }}{4}t,
\end{split}
\end{equation}
\begin{equation}
{C_{r3}}\left( t \right) = - i\frac{{4G}}{{\sqrt {32{G^2} - {\kappa ^2}} }}{e^{ - {{\kappa t} \mathord{\left/
{\vphantom {{\kappa t} 4}} \right.
\kern-\nulldelimiterspace} 4}}}\sin \frac{{\sqrt {32{G^2} - {\kappa ^2}} }}{4}t.
\end{equation}

Same as the large detuning case,  we plot the fidelity $F = \left| {\left\langle {{\psi _{nr}}\left( t \right)\left| {010} \right.} \right\rangle } \right|$ in Fig. \ref{resonant}(a) and possibility $ P = {\left| {{C_{r1}}} \right|^2} + {\left| {{C_{r2}}} \right|^2} + {\left| {{C_{r3}}} \right|^2}$ at $t = \frac{{4\pi }}{{\sqrt {32{G^2} - {\kappa ^2}} }}$ in Fig. \ref{resonant}(b) as a function of $G$ and $\kappa$. Here $\left| {{\psi _{nr}}\left( t \right)} \right\rangle$ is the normalized state of $\left| {\psi_r \left( t \right)} \right\rangle $. It is found that both $P$ and $F$ are in favor of larger $G$ and less $\kappa$. When we choose $\kappa = 75.2$kHz and $G = 50$kHz, the fidelity $F = 0.926$ and the successful possibility $P = 0.59$.
As we can see, for both large detuning and resonant schemes, the quantum state transfer could be realized with pretty high fidelty and successful possibility. In experiment, we can choose either of them for convenience.

\section{conclusion}
In this paper, we propose a scheme to couple librational and translational modes of a levitated nanoparticle with an optical cavity mode. We discuss how to realize quantum state
transfer from a librational mode to a translational mode, and vice versa. We also discuss effects of  cavity decay on the fidelity of state transfer. We find that the high-fidelity
state transfer could be realized under practical experimental conditions.

\section*{Funding and Acknowledgement}
National Natural Science Foundation of China (61435007); Joint
Foundation of Ministry of Education of China; National Science Foundation (NSF) (1555035-PHY).
\\We would like to thank Yue Ma for the helpful discussions.

\end{document}